\documentclass[12pt,preprint]{aastex}

\renewcommand {\deg}   {\mbox{$^\circ$}}

\newcommand   {\kms}   {\mbox{km\,s$^{-1}$}}
\renewcommand {\ga}    {\mbox{\rlap{\hbox{\lower5pt\hbox{$\sim$}}}\hbox{$>$}}}
\renewcommand {\la}    {\mbox{\rlap{\hbox{\lower5pt\hbox{$\sim$}}}\hbox{$<$}}}

\begin{document}
\pagenumbering{arabic} 
\def\kms {\hbox{km{\hskip0.1em}s$^{-1}$}} 
\voffset=-0.8in

\def\msol{\hbox{$\hbox{M}_\odot$}}
\def\lsol{\hbox{$\hbox{L}_\odot$}}
\def\kms{km s$^{-1}$}
\def\Blos{B$_{\rm los}$}
\def\etal   {{\it et al.}}                     
\def\psec           {$.\negthinspace^{s}$}
\def\pasec          {$.\negthinspace^{\prime\prime}$}
\def\pdeg           {$.\kern-.25em ^{^\circ}$}
\def\degree{\ifmmode{^\circ} \else{$^\circ$}\fi}
\def\ut #1 #2 { \, \textrm{#1}^{#2}} 
\def\u #1 { \, \textrm{#1}}          
\def\ddeg   {\hbox{$.\!\!^\circ$}}              
\def\deg    {$^{\circ}$}                        
\def\le     {$\leq$}                            
\def\sec    {$^{\rm s}$}                        
\def\msol   {\hbox{$M_\odot$}}                  
\def\i      {\hbox{\it I}}                      
\def\v      {\hbox{\it V}}                      
\def\dasec  {\hbox{$.\!\!^{\prime\prime}$}}     
\def\asec   {$^{\prime\prime}$}                 
\def\dasec  {\hbox{$.\!\!^{\prime\prime}$}}     
\def\dsec   {\hbox{$.\!\!^{\rm s}$}}            
\def\min    {$^{\rm m}$}                        
\def\hour   {$^{\rm h}$}                        
\def\amin   {$^{\prime}$}                       
\def\lsol{\, \hbox{$\hbox{L}_\odot$}}
\def\sec    {$^{\rm s}$}                        
\def\etal   {{\it et al.}}                     
\def\la{\lower.4ex\hbox{$\;\buildrel <\over{\scriptstyle\sim}\;$}}
\def\ga{\lower.4ex\hbox{$\;\buildrel >\over{\scriptstyle\sim}\;$}}
\def\refitem{\par\noindent\hangindent\parindent}
\oddsidemargin = 0pt \topmargin = 0pt \hoffset = 0mm \voffset = -17mm
\textwidth = 160mm  \textheight = 244mm
\parindent 0pt
\parskip 5pt

\slugcomment{Resubmitted ApJL}
\shorttitle{Sgr A*}
\shortauthors{}

\title{Radio Continuum Emission from the  Magnetar SGR J1745-2900:
Interaction  with Gas Orbiting Sgr A*} 
\author{F. Yusef-Zadeh$^1$, R. Diesing$^1$, M. Wardle$^2$, L. O.  Sjouwerman$^3$, M. Royster$^1$,\\
  W. D. Cotton$^4$,  D. Roberts$^1$ \& C. Heinke$^5$}
\affil{$^1$CIERA, Department of Physics and Astronomy, Northwestern University, Evanston, IL 60208}
\affil{$^2$Department  of Physics and Astronomy,  Macquarie University, Sydney NSW 2109, Australia}
\affil{$^3$National Radio Astronomy Observatory, Socorro, NM 87801}
\affil{$^4$National Radio Astronomy Observatory,  Charlottesville, VA 22903}
\affil{$^5$Department of Physics, University of Alberta, Room \#238 CEB, 11322-89
Avenue, Edmonton AB T6G 2G7, Canada}

\begin{abstract} 

We present radio continuum light curves of the magnetar SGR J1745$-$2900 and Sgr A* obtained with 
multi-frequency, multi-epoch Very Large Array observations between 2012 and 2014. During this period, a 
powerful X-ray outburst from SGR J1745$-$2900 occurred on 2013-04-24. Enhanced radio emission is delayed 
with respect to the X-ray peak by about seven months. In addition, the flux density of the emission from the 
magnetar fluctuates by a factor of 2 to 4 at frequencies between 21 and 41 GHz and its spectral index varies 
erratically. Here we argue that the excess fluctuating emission from the magnetar arises from the 
interaction of a shock generated from the X-ray outburst with the orbiting ionized gas at the Galactic 
center.  In this picture, variable synchrotron emission is produced by ram pressure variations due to 
inhomogeneities in the dense ionized medium of the Sgr A West bar.  
The pulsar with its high transverse 
velocity is moving through a highly blue-shifted ionized medium.  This implies that the magnetar is at a 
projected distance of $\sim0.1$ pc from Sgr A* and that the orbiting ionized gas is partially or largely 
responsible for a large rotation measure detected toward the magnetar. Despite the variability of Sgr A* 
expected to be induced by the passage of the G2 cloud, monitoring data shows a constant flux density and 
spectral index during this period. \end{abstract}


\section{Introduction}

Detection of pulsars orbiting Sgr A*, the massive black hole at the Galactic center, is important because 
they probe the properties of the black hole (Pfahl and Loeb 2004) and the interstellar medium (ISM) of the 
unique environment of the Galactic center. Magnetars are pulsars with high spin down rate and extremely high 
inferred magnetic fields (10$^{14}$ to 10$^{15}$ G) and are generally accompanied by X-ray outbursts.  Only 
four of the 28 known magnetars have exhibited pulsed radio emission (Camilo \etal\, 2006; 
Levin \etal\, 2010). 
The recent 
discovery of the magnetar SGR J1745$-$2900 has opened a new window to study the  ionized medium 
near Sgr A*. The Galactic center magnetar was first discovered by \textit{Swift} as an X-ray outburst on 
JD 2456406 (2013-04-24; Kennea \etal\, 2013). Subsequent X-ray measurements detected 
pulsed emission with a period of 
3.76 s, with a high column density and a small offset from Sgr A* (e.g., Rea \etal\, 2013; Mori \etal\, 2013). 
Radio observations detected pulsed emission 
with a declining spectrum between 4.5 and 20 GHz 
after 
the magnetar went into outburst (Shannon and Johnston 2013; Eatough et al.  2013; Lynch \etal\, 2015; 
Torne \etal\,  2015).  SGR 
J1745$-$2900 has the highest dispersion measure DM=1778 cm$^{-3}$ pc and rotation measure 
RM=$-6.696\times10^4$ rad m$^{-2}$ of any known pulsar in the Galaxy (Shannon and Johnston 2013; Bower 
\etal\, 2014). These radio measurements suggest that SGR J1745$-$2900 is likely to be 
at  a projected distance of 0.1 pc  from Sgr~A*. 
Proper motion measurements show a transverse velocity of 236 \kms\, (Bower 
\etal\, 2015a) which is consistent with the  magnetar being near Sgr A*.

Using the Jansky Very Large Array (VLA), we recently reported brightening of the continuum emission from the magnetar 
SGR J1745$-$2900 at 41 GHz (Yusef-Zadeh \etal\, 2014a). We are motivated to study the light curve of radio emission 
from the magnetar at high frequencies and its temporal correlation with the X-ray emission. 
The magnetar has been experiencing a fast decay in its X-ray flux between 3-10 keV 
followed by a slow decay 200 days after its initial outburst 
(Rea \etal\, 2013; Coti Zelati \etal\, 2015) 
on JD 2456406 (2013-04-24). 
Here,  we  present light curves of radio  emission from the magnetar and
Sgr A*. These radio light curves based on VLA observations 
show  that  the onset of enhanced radio emission from the magnetar  began 
seven months after the date of the X-ray outburst. 
In addition, the light curves of the magnetar indicate that the radio continuum emission at high frequencies 
fluctuates without an obvious pattern on a monthly time scale. 
We discuss the time  delay between  the X-ray outburst and the radio emission 
in terms of the interaction of an outflow  from the magnetar with the material associated with the orbiting 
ionized gas. 

\section{Observations and Data Reduction}

In 2012 October, at the advent of the passage of the G2 cloud with respect to Sgr A* (e.g., Gillessen et al. 2012, 
Narayan et al. 2012),  monitoring of the radio flux density of Sgr A* 
began (see Bower \etal\, (2015b) who use the same NRAO service observations data 
as we do here).
Each observation  lasted for $\sim2$ hours, 
 using up to 2 GHz of continuum bandwidth rotating through the standard 
observing frequencies centered at 1.5, 3.0, 5.5, 9.0, 14.0, 21.2, 32.02 and 41 GHz. 
Typical on-source times per frequency setup were about 6 minutes.
The flux density was calibrated with respect to 3C286 and bandpass shapes were determined using NRAO530
and 3C286. 

The data were calibrated and imaged in AIPS initially and  follow-up analysis was made with CASA. To correct for 
phase instabilities during the scans, phase-only self-calibration was applied (in all observations). Because of the 
compact nature of both Sgr A* and the magnetar, we used extended array configurations (A\&B) and high 
frequencies{\footnote{https://science.nrao.edu/science/service-observing}}. 
The absolute flux 
density of these observations may be inaccurate by up to 10\% as the elevation of the primary flux calibrator in most 
observing runs is far different from that of the Galactic center, introducing errors in modeling the atmospheric 
transmission at the higher frequencies. 
These measurements are consistent with millimeter monitoring data 
indicating a stable flux density of Sgr A* during the same period (Bower \etal\, 2015b). As the measured flux 
densities of the magnetar which lies 2.3$''$ SE of Sgr A*, are tied to the flux density of Sgr A*, this provides an 
excellent platform to perform flux density measurements of the magnetar.

We also present the flux density of the magnetar and Sgr A* from three different epochs of 
observations  on 2011--08--04 (JD2455777), 2014--02--21 (JD 2456710) and 2014-03-09 (JD 2456732). 
Details of these long  observations which were taken in the  A configuration 
are  described in Yusef-Zadeh \etal\, (2014b, 2015). 
 
\section{Results}

As part of the long observations of Sgr A*, we detected a compact source coincident with the magnetar at 34.5 and 44.6 
GHz with a flux density of 1.30$\pm0.01$ and 1.62$\pm0.04$ mJy for an unresolved source on JD 2456732 and 2456710, 
respectively.  The positional offset between SGR 1745-2900 and Sgr A* is 2.3917$''$ and 2.3923$''$  
with statistical errors of 1.9 and 1.1 mas accuracy at 
44.6  and 34 GHz, respectively, and systematic errors of  few mas.  
The comparison of the continuum radio emission from the magnetar with earlier long VLA 
observations taken on JD 2455777 (2011-08-04), with similar resolution and sensitivity, indicate that the magnetar has 
brightened significantly at 42 GHz.  The magnetar was not detected above 3$\sigma\sim243$ $\mu$Jy beam$^{-1}$ at 42 
GHz on JD 2455777. The flux has increased by at least a factor of 7 between these two epochs. Figure 1a, which is 
based on data taken over a long observation on 2014-03-09 (JD 2456732), shows a 34 GHz continuum image of the 
6$''\times3''$ region of the ionized gas surrounding Sgr A* and SGR J1745$-$2900 with a spatial resolution of 0.1$''$.  
The magnetar, as labeled on this figure, is projected to the SE of Sgr A* and at the edge of a shell-like ionized 
feature.  The ionized medium surrounding the magnetar has a high density of 10$^{4-5}$ cm$^{-3}$ and a highly 
blue-shifted velocity --200 \kms\, with broad linewidth (e.g., Roberts \etal\, 1996; Zhao \etal\, 2009). This complex 
region shows a wide range of angular scales including an elongated cavity of ionized gas with surface brightness 
$\sim15\, \mu$Jy beam$^{-1}$ at 34 GHz to the NE of the magnetar extending for $\sim1.5''$.

We used the snapshot monitoring data taken with the VLA at the highest available frequency bands (K, Ka and Q) in the 
most extended configurations (A \& B). 
Figure 1b--d show three images of the medium surrounding the magnetar at 41 
GHz based on snapshot observations taken on JD 2456626, 2456656 and 2456704, respectively. The magnetar is detected 
with a signal-to-noise $\sim$26 in the snapshot observation taken on JD 2456704. The flux densities  of the magnetar and Sgr 
A* are determined in the image plane by fitting a 2D Gaussian to individual sources. The parameters of the fit to both 
long and short observations are listed in Table 1 with entries in columns 1 to 7 corresponding to the observation 
calendar date, the observing date in JD, the observing center frequency, the magnetar and Sgr A* flux densities, the 
VLA array configuration and the synthesized beam, respectively. 
The flux density of Sgr A* is relatively constant with 
a typical flux density $\sim$1.5 Jy at 41 GHz. 

Figure 2a shows the light curves of the magnetar and Sgr A* in blue and red colors, respectively. Because 
the VLA was in its compact configuration (C \& D), the emission from the magnetar during days JD 2456350 and 
JD 2456550 is assumed to be similar to the upper limits given in Figure 2a between JDs 2456250 and 2456550. 
The emission from the magnetar is not detected until JD 2456710 and then fluctuates in all bands by a factor 
of 2 to 4 between epochs. Our B-array data taken at 41 GHz (Fig. 1b,c and the bottom panel of Fig.  2a) show 
3$\sigma$ upper limits of 0.82 (JD 2456592) and 0.70 (JD 2456626) mJy beam$^{-1}$. These values are less 
than the flux density 1.85$\pm0.07$ detected in the A configuration on JD 2456704 at 41 GHz.  The upper 
limit on JD 2456656 is too high to distinguish the magnetar from the diffuse emission. Thus, these 
observations show that the magnetar was not detected at 41 GHz until JD 2456626, seven months after the 
X-ray outburst occurred. It is possible that we did not detect radio emission prior to this period because 
of our poor sensitivity, though our measurements are not inconsistent with the 8.7 GHz monitoring data 
indicating that the flux density of the pulsar became stronger and more erratic after JD 2456726 (Lynch 
\etal\, 2015).

The variable fluxes of the magnetar  and Sgr A* as a function of frequency are 
shown in Figure 2b. Unlike the constant  spectral index   of Sgr A*, 
the magnetar shows an erratic spectral index.
Table 2 provides the values of the spectral index $\alpha$ where the flux density 
is F$_\nu\propto\nu^\alpha$ for three pairs of 
frequencies.  
Entries in column 1 to 5 of Table 2 give the  calendar date, JD, the spectral index 
between 21 and 32 GHz, 32 and 41 and 
21 and 41 GHz, respectively.   The erratic  and steady 
spectral index values of the magnetar and Sgr A*, respectively, 
are noted in column 3, 4 and 5.  
Torne \etal\, 2015) have recently shown  significant flux 
variations on timescales of the order of tens of minutes.  
Since this timescale is not much longer than the on-source time per frequency setup in our data (6 minutes), 
it is possible that there may have been intrinsic evolution in the radio flux 
between observations at different frequencies, making it hard to interpret the measured spectral indices.

Pulsed radio emission from the magnetar SGR J1745$-$2900 was reported to have a flux density of 0.3 mJy beam$^{-1}$ at 
$\nu$=18 GHz with a decreasing flux density at higher frequencies on 2013-05-01 (Shannon \& Johnson 2013). Using the 
spectral index of $\alpha=-1$, the expected flux density of pulsed emission at 42 GHz is 0.12 mJy beam$^{-1}$, 
$\sim7$\% of the flux density detected at 44 GHz, assuming that the magnetar was not showing variable emission.  It 
turned out that monitoring of the magnetar with the GBT indicated that the magnetar underwent a dramatic change in 
flux density and pulse profile morphology at 8.7 GHz near JD 2456713 (Lynch \etal\, 2015). Enhanced 44 GHz emission 
was also detected, as shown in Figure 2a, during this epoch. The GBT observations show also an erratic behavior of 
pulsed radio emission at 8.7 GHz.
More recently, Torne  \etal\, (2015) report the detection of pulsed emission up to 245 GHz.
These measurements are consistent with our high frequency, high resolution 41 GHz continuum data.
It is possible that the  high frequency emission is entirely due to pulsed emission (Torne \etal\, 205).
We have an alternate picture in which  the upulsed emission from the shock  front
could contribute to the total high frequency radio emission.

\section{Discussion \& Conclusion}

The emission from the magnetar is pulsed at radio wavelengths with a nonthermal spectrum. 
Here we propose an 
additional unpulsed component that originates from a shock front and could contribute to radio and X-ray 
emission. We argue that an increase in the continuum flux at high frequencies could in part be explained by 
the magnetar's winds interacting with the ionized gas orbiting around Sgr A*.  The magnetar is moving with a 
high transverse velocity of 236 \kms\, (Bower \etal\, 2015a) with a position angle (PA) $\sim22^\circ$ 
through a highly blue shifted --200 \kms\, ionized medium (Roberts \etal\, 1996; Zhao \etal\, 2009). The 
ionized gas associated with Sgr A West has a transverse velocity that is opposite to that of the magnetar 
(Yusef-Zadeh \etal\, 1998; Zhao \etal\, 2009). This interaction will create a bow shock assuming that the 
magnetar is embedded within the bar of orbiting ionized gas, at a projected distance of 0.1 pc from
 Sgr A*. This would then  strengthen 
the suggestion that the magnetar is at the Galactic center (Bower \etal\, 2015a).

\newcommand{\vrel}{v_\mathrm{rel}}
\newcommand{\nH}{n_\mathrm{H}}

We hypothesize that the magnetar's radio emission is in part produced by synchrotron emission from electrons 
accelerated in a reverse shock arising from the interaction of a pulsar's outflow -- perhaps expanded 
significantly by the magnetar's outburst -- with the dense ionized material in the Sgr A West bar.  To 
estimate the properties of the cometary bubble predicted in this model, we characterize the source region by 
radius $R$, and assume that it is threaded by a uniform magnetic field $B$, and that the relativistic 
electrons between 1\,MeV and 10\,GeV are in equipartition with the magnetic field and have a power-law 
energy spectrum $\propto E^{-1}$, producing a flat synchrotron spectrum.  Then the synchrotron flux is 
\begin{equation}
    S_\nu \approx 0.67 \left(\frac{R}{2\times10^{14}\,\mathrm{cm}}\right)^3\,
    \left(\frac{B}{0.1\,\mathrm{G}}\right)^{7/2}
    \textrm{mJy}
\end{equation}
where we have adopted a distance of 8\,kpc to the Galactic center. 
Equating twice the magnetic pressure in the source, i.e.\ $B^2/4\pi$,
 to the incident ram pressure due to relative motion of the magnetar and 
the surrounding ionized interstellar gas, $P_\mathrm{ram} = \rho  \vrel^2$, we obtain
\begin{equation}
    S_\nu \approx 3.4 \left(\frac{R}{2\times10^{14}\,\mathrm{cm}}\right)^3\,
    \left(\frac{n_H}{10^5\,\mathrm{cm^{-3}}}\right)^{7/2}\,
    \left(\frac{\vrel}{1000\,\mathrm{km\,s^{-1}}}\right)^{7/4}\,
    \textrm{mJy}
    \label{eq:Snu}
\end{equation}

This implies  a source diameter of roughly 20\,AU, equivalent to 2.5\,mas at 8\,kpc.  The pressure in the cometary 
bubble supported 
by the outflow  must also be comparable to the external ram pressure, implying an energy content $E_\mathrm{bubble} = 
(3P_\mathrm{ram})\times (4/3\pi R^3)\approx 2.4\times10^{41}$\,erg for the reference values in equation (2)
The magnetar spin down rate implies $\dot{E} = 1.3\times10^{34}\,$erg\,s$^{-1}$ (Lynch et al.\ 2015) 
so this energy can be replaced on a time scale of about 6 months. Variability of the synchrotron emission may be 
produced by ram pressure variations due to inhomogeneities in the external medium: increased ram pressure compresses 
the cometary outflow bubble ($R\propto P_\mathrm{ram}^{-3/4}$) and increases the field strength in the 
synchrotron source region 
($B\propto P_\mathrm{ram}^{1/2})$; equation (2) 
implies that $S_\nu\propto P_\mathrm{ram}$.
In this picture,  the outflow produces a 20 AU bubble with the  mean expansion speed  of
$\sim$ 170 \kms, seven months after the X-ray outburst  when the magnetar becomes radio loud at high 
frequencies.

This model can also explain the spectral index of the emission from 
the magnetar. For optically thin synchrotron 
radio emission, the energy index p=2 where $\alpha=(1-p)/2$. For different values of the spectral index, the 
spectrum becomes flat or inverted when the emission becomes optically thick. 
Diffusive shock acceleration by a single shock is known to 
produce a power-law energy distribution with $\alpha$=0.5 (Blandford \& Eichler 1987). 
However, multiple, identical and nonidentical shocks can produce
a varying degree of spectral index including a 
flat spectrum (Pope \& Melrose 1994; Melrose \& Pope 1993).
In the context of this model, the synchrotron emission from the bow shock 
should also be polarized at radio wavelengths if the magnetic field is not tangled.
The direction of the magnetic field in the optically this synchrotron source 
is likely to be parallel to the bow shock.
Recent polarization measurements 
show that the intrinsic  PA of  the electric field is $\sim16\pm9^\circ$  (Kravchenko \etal\, 2015),  
which is well within the PA  of the  proper motion of the magnetar (Bower \etal\, 2015a).  
This polarization result is not inconsistent with the expected magnetic field direction in the shock front. 

In addition to nonthermal radio emission, 
the dense interstellar gas is shock-heated to temperatures $\sim 1.4\times 10^7 \vrel^2$\,K as it collides with the 
outflow  bubble, and subsequently emits X-rays while flowing around the obstruction and expanding downstream.  The 
radiative cooling time is about 15 years, so the gas cools primarily by adiabatic expansion on a time scale 
$t_\mathrm{exp} = R/c_s \sim 0.1$\,yr, where $c_s$ is the sound speed in the post-shock gas.  We estimate the 2--8\, 
keV luminosity of the shocked gas using
\begin{equation}
    L_X \approx  \pi (2R)^2 \textstyle{\frac{1}{4}} \vrel t_\mathrm{exp} \times (4\nH)^2 \Lambda(T) \times
    \left[e^{-\mathrm{2\,keV}/kT}-e^{-\mathrm{8\,keV}/kT}\right]
    \label{eqn:Lx}
\end{equation}
where the three factors are the volume of hot shocked gas (determined  using twice 
 the radius of the synchrotron source which itself is 
estimated for a 30\,GHz radio flux of 3\,mJy  using equation (2), 
its  cooling rate per unit volume, and, lastly,  the fraction of 
radiation emitted in the 2--8\,keV window.  
The factors of four account for the compression of the gas as it is shocked.  We adopt an approximation to the 
cooling function computed by B\"oringher and Hensler (1989) for twice solar metallicity: $\Lambda(T) = 1.5 \left( 
[0.4/T_7]^{2.75} + [T_7/0.4]^{0.8} \right) \textrm{ for } 0.2< T_7\leq1 1,\,  \mathrm{or}\,
1.2 \left( [1.5/T_7]^{1.4} + T_7^{0.5} 
\right) \textrm{ for } T_7>1$ in units of $10^{-23}\,  \mathrm{erg\,cm^3\,s^{-1}}$, where $T_7 = T/10^7$\,K.  We plot 
the resulting luminosity in Figure 3
as a function of the assumed relative velocity between the 
magnetar and the dense ionized  material associated with Sgr A West. We note that the flux is three orders of 
magnitude below the current X-ray flux of the magnetar (e.g. Kaspi et al 2014), and so is not currently observable.
However, because of its    quadratic  dependence  on the size of the source, the X-ray luminosity could 
be detected in the future if the emitting region  were ten  times larger. 

The emission measure of the ionized gas in Sgr A West in the vicinity of the magnetar is $n_e^2 L \approx 
2\times10^7$\,cm$^{-6}$\,pc (e.g.~ Roberts \etal\, 1996; Zhao \etal\, 
 2009), yielding $\mathrm{RM}\approx 8\times10^4 \, 
B_\mathrm{0.5mG} / n_5$ \,rad\,m$^{-2}$.  This value   is very similar to the observed value if the  
the  magnetic field is  0.5 mG and the electron density 
$10^5\,\mathrm{cm}^{-3}$. 
Using dust polarization measurements, 
Aitken \etal\,  (1998) estimated a minimum magnetic field of $\sim$2 mG in the bar of
ionized gas.   Our  value of the magnetic field is 
reasonable when compared to those of Aitken et al. (1998).
Thus the ionized orbiting gas is likely responsible for the observed high rotation 
measure, $\mathrm{RM}\approx -6.7\times10^4$\,rad\,m$^{-2}$. Unlike the warm ionized medium within a few tenths of 
Sgr A* proposed here, 
another  recent study of the RM of the 
magnetar suggests that  the  hot X-ray emitting gas within several  parsecs of  Sgr A* is  
responsible for the high RM (Eatough \etal\, 2013). 

In summary, we presented light curves of the magnetar and Sgr A* over the course of 
the monitoring campaign as the G2 cloud approached Sgr A*. 
We find that the  flux density and the spectral index of Sgr A* remained constant whereas
the continuum emission and the spectral index 
of  the magnetar showed an erratic behavior similar to that of its pulsed emission 
(Lynch \etal\, 2015). We noted high frequency 44 GHz emission was detected about seven  months after the 
 X-ray outburst.  We argued that  this time delay  and the enhanced radio 
emission from the magnetar could be 
explained 
by the interaction of the wind from the magnetar with the orbiting ionized gas near Sgr A*. 
This picture can also explain the large RM observed toward the magnetar implying that the 
Faraday screen is localized within  the ionized layer orbiting Sgr A*. 
The continuum radio emission arises from two components,  pulsed and  unpulsed emission. 
The relationship of the pulsed and shocked emission remains unclear.
Future  observations are required to determine  the contribution of shocked emission.

Acknowledgments: We thank the referee for useful comments. 
This work is partially supported by the grant 
AST-1517246 from the NSF. 
The authors wish to thank the NRAO staff for instigating the service monitoring observations and, in particular, 
Claire Chandler for providing the calibrated data and images. The National Radio Astronomy Observatory 
is a facility 
of the National Science Foundation operated under cooperative agreement by Associated Universities, Inc.

\begin{figure}[p]
\centering
\includegraphics[scale=0.75,angle=0]{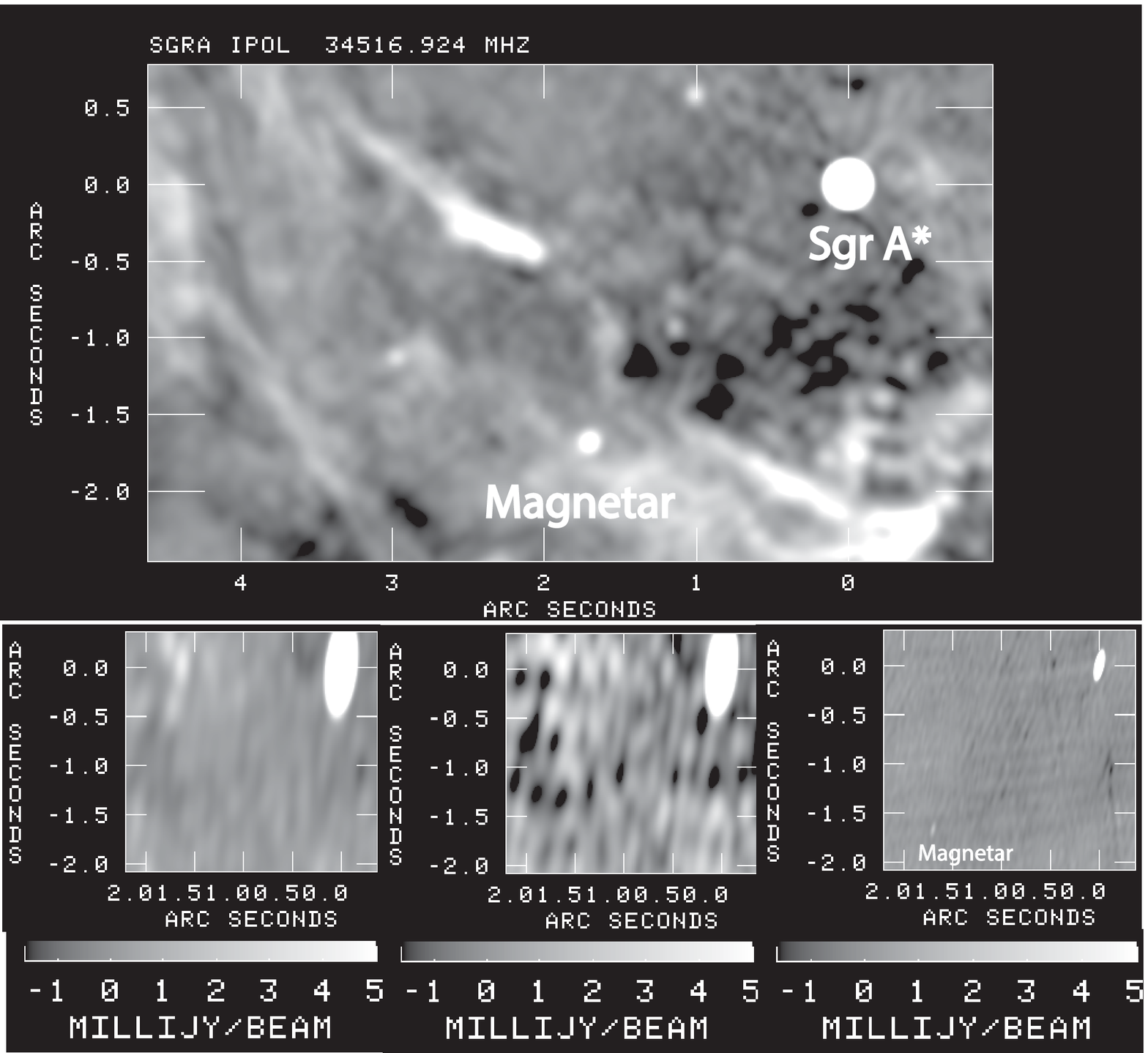}
\caption{
{\it (a) Top}
A grayscale continuum image of the  region surrounding the magnetar SGR J1740-2900
at 34 GHz taken on JD 2456732 (2014-03-09) 
convolved to a resolution of $0.1''$. The grayscale  ranges between 
$-0.13$ and  1 mJy beam$^{-1}$. 
{\it (b--c--d) Bottom} Three 41 GHz images from left to right are based on 
short  observations on JD 2456626 (2013-11-29), 
JD 2456656 (2013-12-29) 
and JD 2456704 (2014-2-15), respectively. 
The right image shows the detection of the magnetar (see Table 1). 
The display range  
is between -1.59 and  5 mJy for all three images. 
}
\end{figure}

\begin{figure}
\centering
\includegraphics[scale=0.45,angle=0]{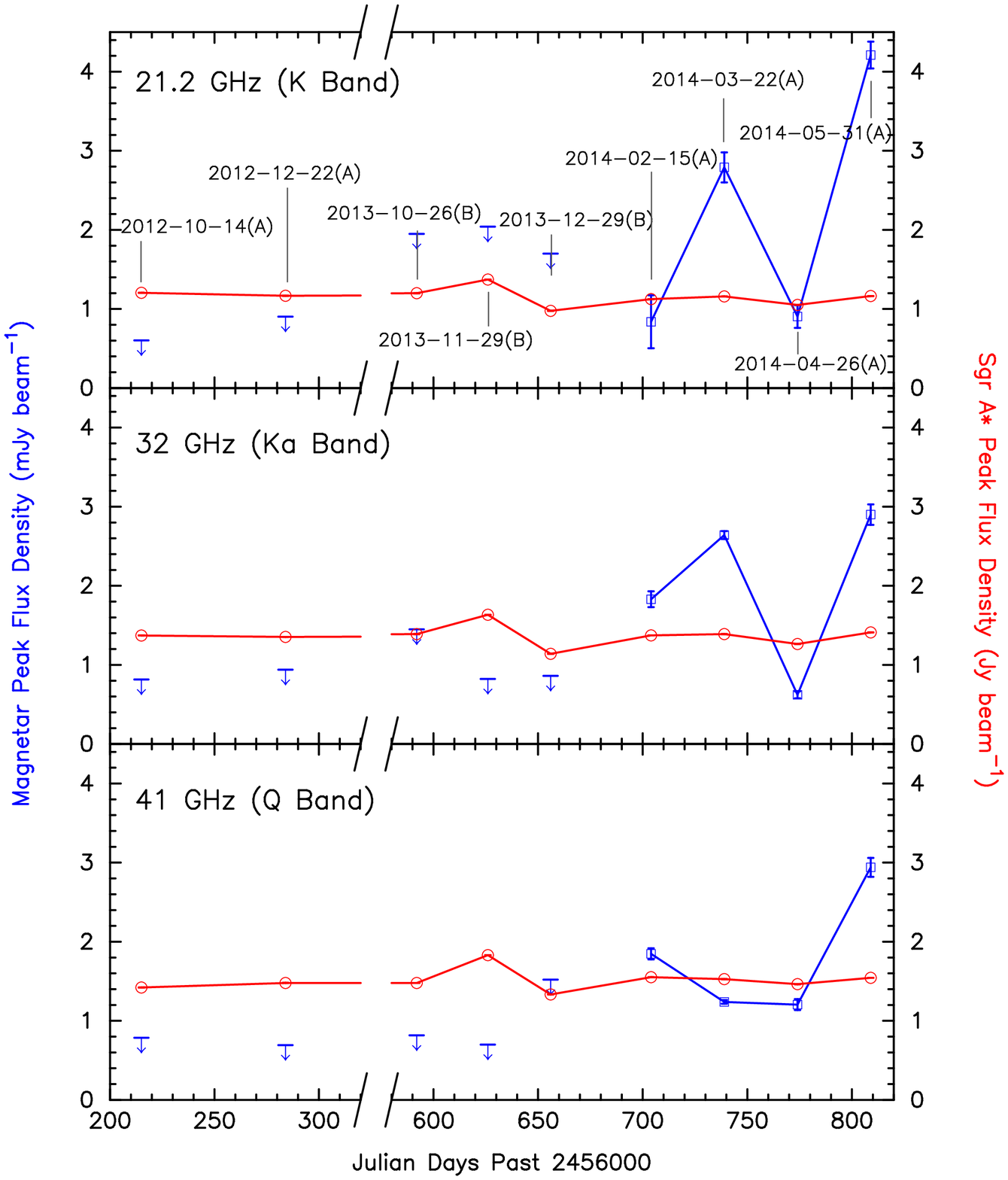}
\includegraphics[scale=0.45,angle=0]{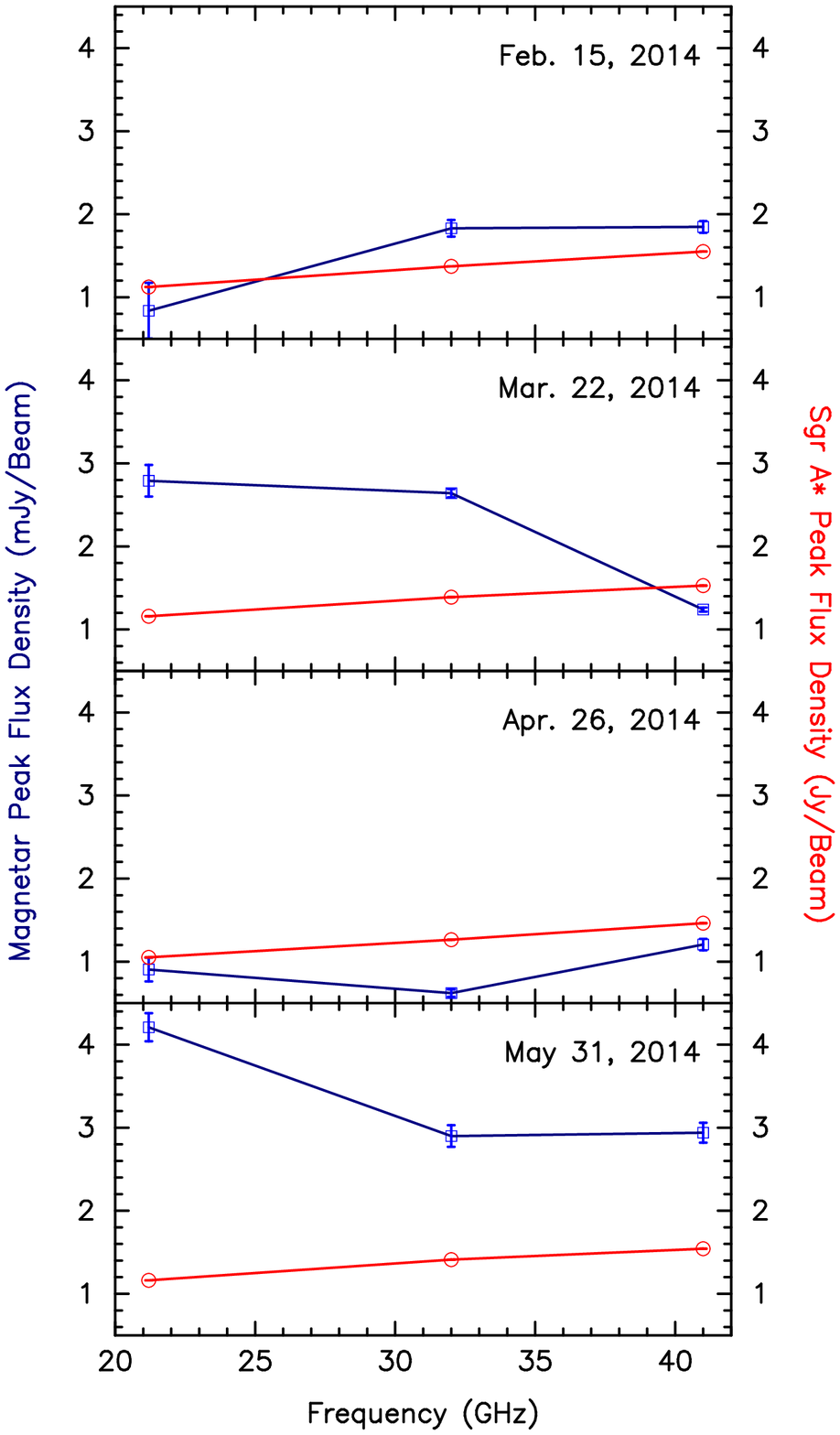}
\caption{
{\it (a) Top}
{Light curves of the magnetar and Sgr A* are shown in blue squares and red circles, respectively. 
Four epochs of observations are plotted at  three  different 
frequencies: 21.2, 32 and 41
GHz, corresponding to K, Ka, Q  bands, 
respectively. The arrows give 3$\sigma$ upper limits of the flux density of 
the magnetar. Note that the flux density scale is in mJy beam$^{-1}$ for the magnetar (blue, left hand side)
and is in Jy beam$^{-1}$ (red, right hand side) for Sgr A*.  The dashed line coincides
with JD 2456406 when the magnetar went into an X-ray outburst. 
Only snapshot observations are used. 
{\it (b) Bottom}
Similar to (a) except that the flux densities  of the magnetar and Sgr A* 
are given as function of  frequency.}
}
\end{figure}

\begin{figure}
\begin{center}
\includegraphics{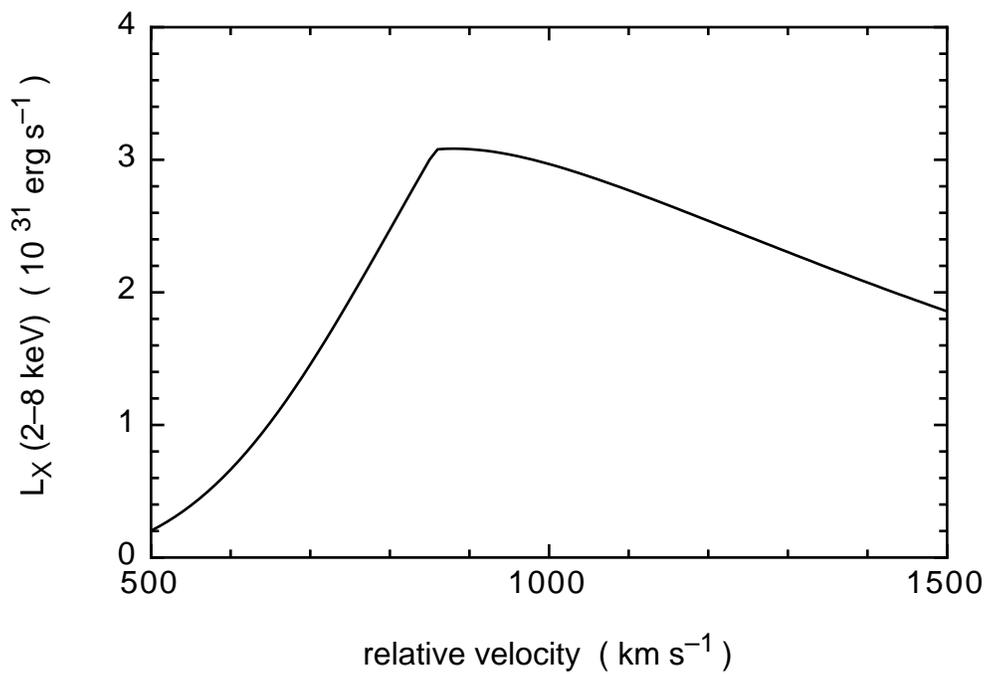}
\caption{Thermal X-ray luminosity produced from the ISM shocked by the relative motion 
of the magnetar and the material in Sgr A West, assuming that the synchrotron radio flux at 30 GHz is 3\,mJy (see text).}
\label{fig:Lx-vb}
\end{center}
\end{figure}

\begin{deluxetable}{lccccccc}
\tablecaption{The Flux Densities of Sgr A* and SGR J1745$-$2900}
\tabletypesize{\scriptsize}
\tablecolumns{8}
\tablewidth{0pt}
\setlength{\tabcolsep}{0.04in}
\tablehead{Date
& \colhead{JD}
& \colhead{Frequency}
& \colhead{Magnetar Flux}
& \colhead{Sgr A* Flux\tablenotemark{b}}
& \colhead{Configuration}
& \colhead{Beam}\\
& 
& (GHz)
& (mJy)
& (Jy)                 
&  (VLA)
& \sl{($\theta_a''\times\theta_b''$(PA$^\circ$))}
}
\startdata                 

2011-08-04&2455777\tablenotemark{c} &  42&   $<$ 0.24\tablenotemark{a}  &       2.18&    A&0.08 $\times$ 0.04 (-5.5)\\       
\hline

2012-10-14 & 2456215 & 21.2  & $<$ 0.60\tablenotemark{a} & 1.20 & A & 0.20$\times$0.06 (5.7)\\
2012-10-14 & 2456215 & 32  & $<$ 0.81\tablenotemark{a}   & 1.37 & A & 0.13 $\times$ 0.04 (-177.6)\\                    
2012-10-14 & 2456215 & 41 &  $<$ 0.78\tablenotemark{a} & 1.42 & A  & 0.11 $\times$ 0.03 (-178.8)\\
\hline
2012-12-22 &  2456284 &  21.2    & $<$ 0.90\tablenotemark{a} &1.17 &A &  0.20 $\times$ 0.06 (9.0)\\
2012-12-22 &  2456284 &  32      & $<$ 0.94\tablenotemark{a} & 1.35 &  A &  0.13 $\times$ 0.05 (-15.5)\\
2012-12-22& 2456284 & 41 & $<$ 0.69\tablenotemark{a} & 1.47&       A&      0.11 $\times$ 0.03  (-13.5)\\
\hline
2013-10-26&       2456592&21.2&   $<$ 1.95\tablenotemark{a}&  1.20& B& 0.66 $\times$ 0.20 (12.7)\\
2013-10-26&2456592&       32&     $<$ 1.45\tablenotemark{a}&  1.39&   B&              0.46 $\times$ 0.13 (9.7)\\        
2013-10-26&2456592&       41&     $<$ 0.82\tablenotemark{a}&  1.48&    B&              0.34 $\times$ 0.10 (7.8)\\        
\hline
2013-11-29&2456626&       21.2&   $<$ 2.04\tablenotemark{a}&  1.37&    B&              0.67 $\times$ 0.19 (-2.1)\\       
2013-11-29&2456626&       32&     $<$ 0.82\tablenotemark{a}&  1.63&    B&              0.44 $\times$ 0.13 (-4.0)\\       
2013-11-29&2456626&       41&     $<$  0.70\tablenotemark{a}&  1.83&    B&              0.34 $\times$ 0.11 (-6.1)\\       
\hline
2013-12-29&2456656&       21.2&   $<$ 1.70\tablenotemark{a}&  0.98&    B&              0.65 $\times$ 0.20 (-2.1)\\        
2013-12-29&2456656&       32&     $<$ 0.861\tablenotemark{a}&  1.14&    B&              0.44 $\times$ 0.13 (-4.2)\\       
2013-12-29&2456656&       41&     $<$ 1.52\tablenotemark{a}&  1.33&    B&              0.34 $\times$ 0.11 (-6.8)\\       
\hline
2014-02-15&2456704&       21.2&   0.84$\pm$0.33&       1.12&A& 0.20 $\times$ 0.06 (-7.6)\\        
2014-02-15&2456704&       32&     1.83$\pm$0.10&       1.37&A& 0.13 $\times$ 0.04 (-9.7)\\        
2014-02-15&2456704&       41&     1.85$\pm$0.07&       1.55&A& 0.11 $\times$ 0.03 (-11.8)        \\ 
\hline
2014-02-21 &2456710\tablenotemark{c} & 44.6&   1.62$\pm$0.04&       1.49&    A&0.07 $\times$ 0.03 (-4.0)\\
2014-03-09&2456732\tablenotemark{c} &  34.5&   1.30$\pm$0.01&       1.72&    A&0.08 $\times$ 0.05 (-1.56)\\             
\hline
2014-03-22&2456739&       21.2&   2.79$\pm$0.19&       1.16&A& 0.20 $\times$ 0.06 (10.6)\\            
2014-03-22&2456739&       32&     2.64$\pm$0.05&       1.39&A& 0.19 $\times$ 0.04 (8.3)\\            
2014-03-22&2456739&       41&     1.24$\pm$0.02&       1.53&A&0.10 $\times$ 0.033 (5.1)\\              
\hline
2014-04-26&2456744&       21.2&   0.90$\pm$0.14&       1.05&A&0.20 $\times$ 0.06 (0.9)\\           
2014-04-26&2456744&       32&     0.62$\pm$0.04&       1.26&A&0.13 $\times$ 0.04 (-1.0)\\          
2014-04-26&2456744&       41&     1.20$\pm$0.07&       1.46&A&0.10 $\times$ 0.03 (-3.5)\\          
\hline
2014-05-31&2456809&       21.2&   4.21$\pm$0.17&       1.16&A&0.23 $\times$ 0.06 (-177.6)\\        
2014-05-31&2456809&32&    2.90$\pm$0.13&       1.41&A&0.15 $\times$ 0.05 (-2.3)\\                  
2014-05-31&2456809&41&    2.94$\pm$0.12&       1.54&A&0.13 $\times$ 0.03 (-4.1)\\                  
\enddata           
\tablenotetext{a}{3$\sigma$ upper limit}
\tablenotetext{b}{1$\sigma$ error is $<0.1$\%}
\tablenotetext{c}{$l$ong observations}

\end{deluxetable}

\begin{deluxetable}{lcccc}
\tablecaption{Spectral Index of SGR J1745$-$2900 and Sgr A*}
\tabletypesize{\scriptsize}
\tablecolumns{5}
\tablewidth{0pt}
\setlength{\tabcolsep}{0.04in}
\tablehead{Date & \colhead{JD} & \colhead{$\alpha1$} & \colhead{$\alpha2$} & \colhead{$\alpha3$}\\
                 &               &   (32--21.2 GHz)   & (41--32 GHz)         &  (41--21.2 GHz)
}
\startdata                 
Magnetar & & & & \\
2014-02-15 & 2456704 & 1.89$\pm$0.95 &	0.04$\pm$0.20 &	1.20$\pm$0.75\\
2014-03-22 & 2456739 & -0.13$\pm$0.17 & -3.05$\pm$0.08 & -1.23$\pm$0.13\\
2014-04-26 & 2456744 & -0.91$\pm$0.41 & 2.67$\pm$0.28  & 0.43$\pm$0.31\\
2014-05-31 & 2456809 & -0.90$\pm$0.14 & 0.05$\pm$0.18 & -0.54$\pm$0.11\\
\\
Sgr A* & & & & \\
2014-02-15 & 2456704 & 0.49$\pm$0.00	  & 0.49$\pm$0.00 & 	0.49$\pm$0.00\\
2014-03-22 & 2456739 & 0.44$\pm$0.00 & 0.38$\pm$0.00  &	0.42$\pm$0.00\\
2014-04-26 & 2456744 & 0.45$\pm$0.00 & 0.59$\pm$0.00  &	0.50$\pm$0.00\\
2014-05-31 & 2456809 & 0.47$\pm$0.00 & 0.36$\pm$0.00 &	0.43$\pm$0.00\\
\\
\enddata           
\end{deluxetable}


\begin{thebibliography}{99}


\bibitem[Aitken et al.(1998)]{1998MNRAS.299..743A} Aitken, D.~K., Smith, 
C.~H., Moore, T.~J.~T., \& Roche, P.~F.\ 1998, \mnras, 299, 743 

\bibitem[Blandford 
\& Eichler(1987)]{1987PhR...154....1B} Blandford, R., \& Eichler, D.\ 1987, \physrep, 154, 1 

\bibitem[Bohriner (1987)]{1989AA} 
B\"ohringer, H. \& Hensler, G. 1989, A\&A, 215, 147


\bibitem[Bower et al.(2014)]{2014ApJ...780L...2B} Bower, G.~C., Deller, A., 
Demorest, P., et al.\ 2014, \apjl, 780, L2 


\bibitem[Bower et al.(2015a)]{2015ApJ...798..120B} Bower, G.~C., Deller, A., 
Demorest, P., et al.\ 2015a, \apj, 798, 120 


\bibitem[Bower et al.(2015b)]{2015ApJ...802...69B} Bower, G.~C., Markoff, 
S., Dexter, J., et al.\ 2015b, \apj, 802, 69 


\bibitem[Camilo et al.(2006)]{2006Natur.442..892C} Camilo, F., Ransom, 
S.~M., Halpern, J.~P., et al.\ 2006, \nat, 442, 892 


\bibitem[Coti Zelati et al.(2015)]{2015MNRAS.449.2685C} Coti Zelati, F., 
Rea, N., Papitto, A., et al.\ 2015, \mnras, 449, 2685 




\bibitem[Gillessen et al.(2012)]{2012Natur.481...51G} Gillessen, S., 
Genzel, R., Fritz, T.~K., et al.\ 2012, \nat, 481, 51 


\bibitem[Eatough et al.(2013)]{2013Natur.501..391E} Eatough, R.~P., Falcke, 
H., Karuppusamy, R., et al.\ 2013, \nat, 501, 391 


\bibitem[Kennea et al.(2013)]{2013ApJ...770L..24K} Kennea, J.~A., Burrows, 
D.~N., Kouveliotou, C., et al.\ 2013, \apjl, 770, L24 




\bibitem[Levin et al.(2010)]{2010ApJ...721L..33L} Levin, L., Bailes, M., 
Bates, S., et al.\ 2010, \apjl, 721, L33 


\bibitem[Lynch et al.(2015)]{2015ApJ...806..266L} Lynch, R.~S., Archibald, 
R.~F., Kaspi, V.~M., \& Scholz, P.\ 2015, \apj, 806, 266 




\bibitem[Melrose 
\& Pope(1993)]{1993PASAu..10..222M} Melrose, D.~B., \& Pope, M.~H.\ 1993, Proceedings of the Astronomical Society of Australia, 10, 222 



\bibitem[Mori et al.(2013)]{2013ApJ...770L..23M} Mori, K., Gotthelf, E.~V., 
Zhang, S., et al.\ 2013, \apjl, 770, L23 




\bibitem[Narayan et al.(2012)]{2012ApJ...757L..20N} Narayan, R., {\"O}zel, 
F., \& Sironi, L.\ 2012, \apjl, 757, L20 


\bibitem[Pfahl 
\& Loeb(2004)]{2004ApJ...615..253P} Pfahl, E., \& Loeb, A.\ 2004, \apj, 615, 253 

\bibitem[Pope 
\& Melrose(1994)]{1994PASAu..11..175P} Pope, M.~H., \& Melrose, D.~B.\ 1994, Proceedings of the Astronomical Society of Australia, 11, 175 

\bibitem[Rea et al.(2013)]{2013ApJ...775L..34R} Rea, N., Esposito, P., 
Pons, J.~A., et al.\ 2013, \apjl, 775, L34 



\bibitem[Roberts et al.(1996)]{1996ApJ...459..627R} Roberts, D.~A., 
Yusef-Zadeh, F., \& Goss, W.~M.\ 1996, \apj, 459, 627 


\bibitem[Shannon 
\& Johnston(2013)]{2013MNRAS.435L..29S} Shannon, R.~M., \& Johnston, S.\ 2013, \mnras, 435, L29 


\bibitem[Torne et al.(2015)]{2015MNRAS.451L..50T} Torne, P., Eatough, 
R.~P., Karuppusamy, R., et al.\ 2015, \mnras, 451, L50 



\bibitem[Yusef-Zadeh et al.(2014a)]{2014ATel.6041....1Y} Yusef-Zadeh, F., 
Roberts, D., Heinke, C., et al.\ 2014, The Astronomer's Telegram, 6041, 1 


\bibitem[Yusef-Zadeh et al.(1998)]{1998ApJ...499L.159Y} Yusef-Zadeh, F., 
Roberts, D.~A., \& Biretta, J.\ 1998, \apjl, 499, L159 


\bibitem[Yusef-Zadeh et al.(2014b)]{2014ApJ...792L...1Y} Yusef-Zadeh, F., 
Roberts, D.~A., Bushouse, H., et al.\ 2014, \apjl, 792, L1 


\bibitem[Yusef-Zadeh et al.(2015)]{2015ApJ...801L..26Y} Yusef-Zadeh, F., 
Roberts, D.~A., Wardle, M., et al.\ 2015, \apjl, 801, L26 


\bibitem[Zhao et al.(2009)]{2009ApJ...699..186Z} Zhao, J.-H., Morris, 
M.~R., Goss, W.~M., \& An, T.\ 2009, \apj, 699, 186 


\end{thebibliography}
\end{document}